# Adaptive Variable Degree-*k* Zero-Trees for Re-Encoding of Perceptually Quantized Wavelet-Packet Transformed Audio and High Quality Speech

Omid Ghahabi [a] and Mohammad H. Savoji [b,] *

[a, b] *Department of Electrical and Computer Engineering, Shahid Beheshti University, Evin Sq., Tehran 1983963113, Iran*
[a] *Now with Research Center of Intelligent Signal Processing (RCISP), Tehran, Iran*
E-mail Address: [a] ghahabi@rcisp.ac.ir, [b] m-savoji@sbu.ac.ir
* Corresponding author. Tel: +98 21 29904162; Fax: +98 21 22431804.

**ABSTRACT**

A fast, efficient and scalable algorithm is proposed, in this paper, for re-encoding of perceptually quantized wavelet-packet transform (WPT) coefficients of audio and high quality speech and is called "adaptive variable degree-*k* zero-trees" (AVDZ). The quantization process is carried out by taking into account some basic perceptual considerations, and achieves good subjective quality with low complexity. The performance of the proposed AVDZ algorithm is compared with two other zero-tree-based schemes comprising: 1- Embedded Zero-tree Wavelet (EZW) and 2- The set partitioning in hierarchical trees (SPIHT). Since EZW and SPIHT are designed for image compression, some modifications are incorporated in these schemes for their better matching to audio signals. It is shown that the proposed modifications can improve their performance by about 15-25%. Furthermore, it is concluded that the proposed AVDZ algorithm outperforms these modified versions in terms of both output average bit-rates and computation times.

## 1. Introduction

Most speech coding systems in use today are based on telephony narrowband speech, nominally limited between 200 and 3400 Hz and sampled at 8 kHz sampling frequency. This bandwidth limitation affects adversely the communication quality. In wideband speech coding, the signal is band-limited to 50-7000 Hz and sampled at 16 kHz resulting in close to face-to-face speech quality. Compared to narrowband telephony speech, the low-frequency extension from 200 to 50 Hz contributes to increased naturalness, presence and comfort. The high-frequency extension from 3400 to 7000 Hz provides better fricative differentiation and therefore higher intelligibility. These increases in bandwidth add also a feeling of transparent communication and eases speaker recognition [1].

Speech coding or compression is then of paramount importance in a viable application of wideband speech communication, when each sample is represented by 16-bit integers, as the resulting raw data amounts to 256 kbit/sec.

In this paper, we introduce an audio and high quality speech coder composed of three main parts: wavelet packet transform (WPT), perceptual quantizer and re-encoder. Although our main focus is on the re-encoder, some novelties, nevertheless, are also introduced in other parts.

In most transform coders, the re-encoder has an important role in the reduction of the output bit-rate. Two re-encoders commonly used are the Huffman and Arithmetic entropy re-encoders. However, there are two main drawbacks namely: disability of generating fixed bit-rates easily, and using more computation time, mostly more than double, in decoding than coding (especially in the case of Arithmetic coding); what is not suitable in most applications. Therefore, designing an efficient re-encoder that does not suffer from these problems is considered worthwhile and is pursued in this paper.

The input signal to the codec, used in this research work, is a 7 kHz speech or music sampled at 16 kHz sampling frequency and represented with 16 bit/sample. All signals are normalized between -1 and 1 prior to their use. Therefore, the input signal is either audio or wideband high quality speech. All the results reported here are obtained by overlapping input frames of 512 samples. The reconstructed signals are assessed in terms of both objective and subjective measures. Subjective qualities are objectively evaluated employing PESQ (Perceptual Evaluation of Speech Quality) scores. The PESQ scores are computed using the source code available from the CD-ROM included in [2] whose results are shown to have high correlations with subjective mean opinion scores (MOS). It is to be noted that the used PESQ algorithm works both for 8 and 16 kHz sampling frequencies and is, therefore, adequate for evaluation of our wideband coded speech files. Furthermore, the objective qualities are evaluated by the segmental signal-to-noise ratio (SEGSNR) results. The average SEGSNR is calculated for 10-msec segments according to equation (1).

$$SEGSNR = \frac{10}{L}\sum_{i=0}^{L-1} log_{10}\left\{\frac{\sum_{n=0}^{N-1} s^2(iN+n)}{\sum_{n=0}^{N-1}\left(s(iN+n) - \hat{s}(iN+n)\right)^2}\right\}, \qquad (1)$$

where $s(n)$ and $\hat{s}(n)$ are the original and the reconstructed signals, respectively. $N$ is the length of each segment (in samples) and $L$ represents the number of the segments.

The paper is organized as follows: Section 2 describes the three main parts of the encoder and addresses the three different re-encoders used including the proposed Adaptive Variable Degree-$k$ Zero-tree (AVDZ). Section 3 compares the re-encoder performances in terms of average bit-rates and comparative computation times. Section 4 concludes and points to some directions for future work.

## 2. General description of used codec

The functional block diagram of this codec is shown in Fig. 1. The encoder operates on frame by frame basis. The first part of the encoder is the wavelet packet transform (WPT) which maps the input frame from the time domain onto a transform space where the signal can be represented with a limited number of high energy coefficients and a high number of low values near zero [3,4]. The next block is the perceptual quantizer which converts the real value output of the transform block into integers based on some perceptual considerations. The output of the quantizer is input to the re-encoder block. No information is lost in this part in the variable bit-rate setup. However, the proposed AVDZ algorithm and the other two zero-tree based ones, being embedded, have the possibility of being used in a fixed bit-rate setup where obviously some information loss is encountered. The re-encoder, in the block diagram, is one of the three schemes compared in the following. The output of the re-encoder is packed with some side information and is saved in a buffer either to be read out at the input of a fixed capacity channel, or packetized for packet switched networks, or to be saved in a file on a computer for instance. The decoder, as shown in Fig. 1, reconstructs the signal by reversing the operations that have taken place in the encoder.

### 2.1. Wavelet packet transform

Wavelet packet (WP) is chosen as transform here for two main reasons: 1- This transform has all properties of wavelet transform (WT) for effective descriptions of natural signals with non-stationarities, 2- In order to benefit from psycho-acoustics of the hearing critical bands, the WP is more flexibly expandable by means of continuing the outputs of the expansion tree; in contrast to dyadic WT with a fixed expansion possibility [5]. Figure 2 shows how the signal bandwidth is decomposed to 19 relatively critical sub-bands in the region of 0-7 kHz. The difference between our method and other similar ones is the combination of WP outputs in different levels of wavelet expansion tree wherever a perfect match with the critical bands is not possible (sub-bands 14-19). These 19 critical sub-bands are summarized in Table 1. Moreover, we have used the new wavelet kernel exploited in [6]. The transfer function of this mother kernel is a parametric and explicit formula based on the well-known raised cosine function:

$$T(f) = \begin{cases} 1, & |f| \leq (r-\beta) \times k, \\ \cos\left[\frac{\pi}{4\beta}\left(\frac{f}{k} - r + \beta\right)\right], & |f| \leq (r+\beta) \times k, \\ 0, & |f| > (r+\beta) \times k, \end{cases} \quad (2)$$

where $f = 0, \ldots, k$ is the frequency and $k = N/2$ with $N$ being the filter order. Furthermore, $r = 1/2$ to have a half-band filter and $\beta$ defines the sharpness of the cut-off or the bandwidth of the overlap between the low and high frequency half-bands. The FIR coefficients are calculated using IFFT of the above expression. These kernels of different $N$ lengths are coined "sav$N$" [6]. These kernel coefficients, once normalized, constitute the low-pass filter component of the filter bank. Using orthogonal relations, the corresponding high frequency half-band analyzing filter and, the low and high half-band synthesizing filters are calculated [7]. Figure 3 shows the mother kernel and the scaling function used in this paper for order $N = 8$ or "sav8" and $\beta = 0.168$. It has been shown that this kernel can achieve better compression for the same filter order $N \geq 8$ in comparison to other well-known kernels such as "db$N$" or Daubechies $N$ [6].

## 2.2. Perceptual quantizer

Using the psychoacoustic properties of the human ear plays an important role, in bit-rate reduction, in most transform codecs. These properties help to distribute the quantization noise in the least sensitive regions of the spectrum, so that its perceptual effect is minimized. Two of the most important and usable psychoacoustic models are those employed in layers 1 to 3 of MPEG-1 standard [8-12]. But, these models are very complex and their uses are time-consuming due to the involved computation in the Fourier spectral domain and complex bit-allocation processes. Furthermore, these models are not compatible with WPT coefficients. Although some research works have attempted to solve these problems [13-18], the use of these models remains still computationally fairly expensive. Therefore, looking for a method that, without using such psychoacoustic models, could reach the desirable bit-rates and perceptual qualities by only taking into account some perceptual considerations is of interest.

In effect, we would like to design a perceptual quantizer that in addition to simplicity has the ability of decreasing the output bit-rates without any noticeable losses in the perceptual qualities. In our heuristic approach we proceed as follows once the input frame is decomposed into critical sub-bands of Table 1:

a) By definition, the narrow-band speech signal is band limited to 300-3400 Hz and 200-3200 Hz in Europe and USA, respectively [11]. However, this bandwidth is considered to be 200-3400 Hz, in general [1]. So, we classify the sub-bands of Table 1 into two main categories: 1- narrow sub-bands (sub-bands 3-16 corresponds to 250-3750 Hz), and 2- wide sub-bands (sub-bands 1-2 and 17-19).

b) We assign an Importance Factor (called I-Factor) to each sub-band in terms of which the sub-bands are allocated 6 or 7 bits, including the sign bit (for the average bit-rates around 32 kbit/sec). In fact, the sub-band with the largest I-Factor is allocated 7 and, that with the smallest is assigned 6 bits. And, the other sub-bands are classified to 7-bit or 6-bit categories according to the closeness of their I-Factors to these maximum and minimum values. The I-Factor assigned to each sub-band is defined in terms of its average energy ($\bar{E}_i$) as in:

$$I - Factor_i = \bar{E}_i^{\alpha}, \qquad \bar{E}_i = \frac{\sum_{j=1}^{L_i} |C_{i,j}|^2}{L_i}, \qquad 1 \leq i \leq 19, \tag{3}$$

where $L_i$ is the length of the sub-band $i$ (in terms of the number of the coefficients), $C_{i,j}$ is the $j^{th}$ coefficient in that sub-band, and $\alpha$ is a factor that controls the bit-rate and the perceptual quality. We can observe an increase in the average bit-rate and the perceptual quality (in PESQ) by decreasing $\alpha$ in the plots of the variations of these values for different audio and speech signals (Fig. 4). By selecting the factor $\alpha$, we have freedom to reach, finely, the desirable bit-rate and perceptual quality. Since high perceptual quality is sought, we select $\alpha = 0.04$.

c) As seen in Fig. 5, since the human ear is less sensitive in wideband regions (sub-bands 1-2, 17-19), in comparison to their adjacent narrow-bands, we reduce the 6 or 7 allocated bits previously in the wide sub-bands. These reductions are experimentally set to 1 and 2 bits for sub-bands 2 and 1, and 1, 3 and 2 bits for sub-bands 17-19, respectively. Our experimental results show that allocating more bits to these sub-bands increases highly the output bit-rate without any considerable increase in the resulting perceptual quality.

d) When the average energy of the input frame is insignificant so is the average energy of the corresponding WPT coefficients. As a re-encoder is used in our codec, employing a scaling factor was decided against. In fact, its impact was investigated experimentally and it was shown that the advantages of its utilization were overshadowed by its negative effects. Since no scale factors are used to scale up these coefficients, they remain small and the number of the allocated bits previously may not be sufficient to avoid equaling them to zero. Our experiments show that nullifying the small coefficients of narrow sub-bands is perceptible when the total energy of the narrow-band region is small. To avoid this, we act as follows:

We calculate the average energy of the coefficients of the narrow-band region (sub-bands 3-16). If this is smaller than a given value $e_1$, we add one bit to the number of the allocated bits of the narrow sub-bands whose average energies are smaller than a defined value $e_2$. In other words, we quantize more accurately these coefficients. To determine the values $e_1$ and $e_2$ we first assume a small arbitrary value for $e_1$ and then determine $e_2$ in the way $\alpha$ was determined in (b). That is, we plot the diagrams of the average bit-rates and their corresponding perceptual qualities (in PESQ scores) versus $e_2$ for a randomly selected signal, and using these curves we select a value $e_2$ for which the relative increase in the bit-rate is justified with a higher increase in quality. The selected $e_2$ for an arbitrary small $e_1$ is then fixed and $e_1$ is determined similar to $e_2$ i.e., the same curves are plotted this time for varying $e_1$ and the same criterion is used to select $e_1$ for the given $e_2$. This procedure is repeated until they remain fairly constant. The chosen parameters are then used in a similar way on a number of audio and speech signals and finally the pair of $e_1$ and $e_2$ that are judged adequate for all signals are selected. Here, $e_1$ and $e_2$ are set to 0.01 and 0.004, respectively.

e) Our studies show that for input frames whose narrow-band region's average energies are larger than a specific limit, for instance $e_3$, the bit allocations can be reduced, by one bit without any noticeable losses in perceptual qualities, in their narrow sub-bands if the average energies of these sub-bands are larger than a defined value $e_4$. Computing $e_3$ and $e_4$ is similar to that of $e_1$ and $e_2$ explained previously. In our work, $e_3$ and $e_4$ are experimentally computed as 0.02 and 0.04, respectively.

Figure 6 shows the range of final allocated bits for each sub-band. As seen in this Figure, the number of the allocated bits to each wide sub-band can be one of 2 possibilities while those assigned to narrow sub-bands are one of 4 cases. So, the used case can be identified in the decoder by sending correspondingly 1 and 2 bit codes as side information. In the result section 3, we will see that this perceptual quantizer based on steps 1 to 5 can reduce the output average bit-rates by about 15-35% (with almost the same perceptual quality or even better) in comparison to all coefficients quantized with a fixed number of bits (7 bits in this case).

*2.3. Re-encoder*

As said previously, the re-encoder usually plays an important role in bit-rate reduction in most transform audio, image or video codecs. Two common re-encoders used are usually the Huffman and Arithmetic entropy re-encoders, or their combinations with Run Length coders. But these re-encoders have two main drawbacks: 1- disability of generating fixed bit-rates easily, and 2- using more computation time in decoding than in coding. These drawbacks make these re-encoders unsuitable for most applications. Some re-encoding schemes do not suffer from these drawbacks. For instance, EZW [19] and SPIHT [20] are two well-known algorithms of this sort. Since these algorithms are meant principally for image coding, they need some modifications for their better matching to audio signals. In [21], their implementations for perceptually audio and high quality speech coding are evaluated, and some modifications are proposed (although a different perceptual quantizer is used). Also, in [22] the performances of these two re-encoders are compared with two re-encoders based on the JPEG's entropy/run length coding. The latter paper concludes that the zero-tree-based algorithms such as EZW and SPIHT that do not suffer from the above mentioned problems have good potentials to replace the entropy re-encoders. In this sub-section we first overview the EZW and SPIHT algorithms and their modified versions (using the above explained perceptual quantizer), then propose a new scheme coined "adaptive variable degree-$k$ zero-tree (AVDZ) algorithm" which is superior to these modified versions both in output average bit-rates and comparative computation times.

### 2.3.1. EZW and Modified EZW

One of the beneficial properties of the wavelet transform, relative to data compression, is that it tends to compact the energy of the input into a relatively small number of wavelet coefficients [23]. It has been recognized that low bit-rate, low mean squared error (MSE) coders can be achieved by coding only the relatively few high energy coefficients [24]. The only problem with this idea is that since only a selected number of coefficients are now being coded, the coder needs to send position information as well as magnitude information for each of the coefficients so that data can be decoded properly. Depending on the method used, the amount of resources required to code the position information can be a significant fraction of the total, negating much of the benefit of the energy compaction. Various ways of lowering the cost of coding the position information associated with the significant coefficients have been proposed. One that achieves a low bit-rate and high quality image coding is named EZW [19]. The EZW algorithm defines this position information as significance maps. A significance map indicates whether a particular coefficient is zero or nonzero (i.e., significant) relative to a given threshold. The EZW determines a very efficient way to code significant maps not by coding the location of the significant coefficients, but rather by coding the location of zeros. It hypothesizes that if a coefficient in a wavelet transform, named parent, is insignificant with respect to a given threshold $T_\ell$, then all wavelet coefficients of the same orientation in the same spatial location at higher frequency sub-bands, named descendants, are likely to be insignificant. The EZW called these set of insignificant coefficients zero-trees [25]. Zero-trees are very efficient for coding since by declaring only one coefficient as a zero-tree root, a large number of descendant coefficients are automatically known to be zero and none of them is coded in that level. In each level, the threshold is half of the preceding one $(T_{\ell+1} = T_\ell/2, 1 \leq \ell \leq last\ step\ level)$. The initial threshold $T_1$ is determined as the largest number power of 2 which is smaller than the maximum of the absolute value (magnitude) of all coefficients. The *last step level* is the level with threshold 1 if the coefficients are quantized and converted to the integer values or obtained on the basis of the target bit-rate.

The algorithm consists of two main passes named Dominant and Subordinate to which two main lists with the same names are associated in each level. All coefficients are placed, first, in the Dominant list and are scanned during the Dominant pass in a predetermined order, and are classified into four types: POS, NEG, IZ and ZTR according to the current threshold $T_\ell$. POS and NEG represent a positive and negative significant coefficient, respectively. IZ stands for isolated zero which means itself is insignificant while one or more of its descendants are significant. ZTR is a zero-tree root and indicates that itself and all its descendants are insignificant. When POS or NEG occurs, its magnitude is put in the subordinate list. Decoder sets its amplitude to $1.5 \times T_\ell$ or $-1.5 \times T_\ell$ while receiving POS or NEG, respectively. The two most important bits (included the sign bit) of binary representation of each significant coefficient are called the dominant bits and the others are called the subordinate ones [19]. Subordinate bits are gradually measured and encoded during the Subordinate passes. In fact, these passes determine the value of the next most significant bit of the binary representation of coefficients on the Subordinate list. This is equivalent to finding the quantized value of these coefficients in comparison to the quantization step size $T_\ell/2$. For example, after the first Dominant pass, the coefficients in the Subordinate list lie in the interval $[T_1, 2T_1)$. The Subordinate pass outputs "1" if a coefficient lies in the upper half of this interval or, "0" if it is in the lower half. Decoder adds to, or subtracts from, the reconstructed coefficient the amount $T_\ell/4$ when receiving the refining bit of "1" or "0", respectively. The threshold is halved and all operations are repeated until the algorithm reaches the *last step level*. In fact, the EZW algorithm sorts the bit order of coded bits so that the most significant bits are sent first. An important end result of the most significant bits being sent first is that the coded bit stream is embedded. This means that bits needed to represent a higher fidelity image can be derived by simply adding extra refining bits to the lower fidelity image representation. Equivalently, a lower fidelity image can be derived by simply truncating the embedded bit stream, resulting in a lower overall bit-rate. More details can be found in [19,25].

*Modified EZW*

The original method uses dyadic WT for image coding and exploits the spatial frequency map to represent the WT coefficients. Similarly, we use the time-frequency map of Fig. 7 for audio signals. This figure shows the WPT coefficients of an exemplary 256-sample input frame. It is noted that the sub-bands 14-19 in Table 1, which are the combination of some other sub-bands, must be separated before being located in the time-frequency map. So, the 19 sub-bands are converted to 26 ones. Each sub-band corresponds to a different frequency space representation but, of the same time duration as the input frame. Each coefficient is located in an uncertainty box that its area is the same for all coefficients on the basis of Heisenberg's uncertainty principle ($\Delta t \times \Delta f = const.$). Furthermore, the coefficients are represented as $c(i,j)$ or $c_{i,j}$, $1 \leq i \leq 26, 1 \leq j \leq L_i$ and $L_i$ is the length (the number of the coefficients) of the sub-band $i$. In other words, $c(i,j)$ is defined as the $j^{th}$ coefficient in the $i^{th}$ sub-band, two examples are shown in Fig. 7. In addition, a coefficient is considered as a descendant of another coefficient (parent as defined below), in the lower frequency sub-band (situated higher up in Fig. 7), if its uncertainty time span is the subset of the time span of that parent. For example, $c(22,2)$ is and $c(17,1)$ is not considered as descendant of $c(9,1)$. The coefficients the descendants refer to are named parents and, direct or immediate descendants are named children.

However, exploiting the EZW and other similar algorithms for perceptual audio signal compression meets the following problems:

a) The statistical assumptions considered in the original EZW may not be valid for WPT coefficients of audio signals.
b) Allocating different number of bits to quantize the coefficients of different sub-bands may weaken the statistical assumptions of EZW more.
c) As seen in Fig. 7, the lengths of some sub-bands are smaller than those of the preceding ones and hence they cannot be considered as their children.
d) EZW and other similar algorithms have their best efficiency when a parent coefficient has more energy than its descendants. This way, if the parent magnitude is lower than the used threshold at that level, it is guaranteed that the parent is a zero-tree root and none of its descendants is coded at that level. But, in practice it is not always so. As seen in the example shown in Fig. 8, the signal energy is not always compressed in low frequency sub-bands and does not show any special order. In fact, the high energy coefficients of upper sub-bands avoid coding the coefficients in lower sub-bands as a ZTR. This reduces extremely the efficiency of EZW and other zero-tree-based algorithms.

The problem (c) is solved for perceptually image compression in [26,27]. This solution is equivalent to change the frequency locations of sub-bands, in the time-frequency map of Fig. 7, for audio signals so that no sub-band has shorter length than the preceding one. This is a way to achieve only full tree structure and does not improve the efficiency problem mentioned in (d). This problem is pointed out in [28]. In that paper, an audio frame is partitioned into 8 (linearly) or 10 (logarithmically) different segments for which a separate sub-tree is selected from four different types of trees that are designed for ascending, descending, concave, and convex coefficient-magnitude behavior within a segment. This way, on the one hand, the problem

(d) is partly solved but, on the other hand, the computation cost is increased because partitioning each frame into 8 or 10 segments increases the transform computation costs by 7 or 9 times. Up to now, few EZW-based research studies have dealt with audio signals. For instance in [29], once a coefficient is found significant, its binary representation is transmitted to the decoder, using perceptual considerations and, there is no need to transmit the refinement bits. In another work [30], it is empirically assessed that if one coefficient is insignificant, there is a relatively high probability that the coefficients of its harmonics are also insignificant. Thereby, four harmonic sub-trees are formed and a relative improvement in the bit-rate is achieved. However, except in [28] none of the above modified schemes address the problem mentioned in (d). As shown in Fig. 8, we can solve this problem, to a great extent, if we adaptively order the sub-bands according to one of the three ways: 1- the average energy $\left(\sum_{j=1}^{L_i}|c_{i,j}|^2/L_i\right)$, 2- the average magnitude $\left(\sum_{j=1}^{L_i}|c_{i,j}|/L_i\right)$, and 3- the maximum magnitude of their coefficients. As mentioned previously, $L_i$ indicates the number of the coefficients (length) in sub-band $i$. This way, we can increase the probability that the parent coefficient magnitude be higher than its descendant ones. Our experimental results show that ordering the sub-bands according to their average magnitude (2- above) yields the best efficiency. For doing this, we must transmit the ordering sequence as side information. However, we may not need to transmit all such sequences because in those sub-bands where all coefficients are zero (e.g., upper bands) the average magnitude is also zero and there is no need for ordering. For example, suppose for instance that for a given frame there are only four sub-bands for which the coefficients average magnitudes are non-zero. The sub-band numbers are assumed to be 2, 5, 7, and 8 ordered as: 5, 8, 7, and 2. So, after ordering the sequence of ordered sub-bands becomes 5, 8, 7, 2, 1, 3, 4, 6, 9, …, 26. Since there is no ordering priority after sub-band 2, the coder sends only the first four numbers by 5-bit codes in additions to a 5-bit zero code indicating the end of the ordering sequence. Although ordering is carried out at some extra cost but, it results in an overall bit reduction as seen later. Furthermore, by doing so, the coefficients with higher magnitudes are sent earlier than that is usually done in EZW with some beneficial effect in bit stream truncation. It is noted that, it is not possible to combine this scheme and that used in [26,27] (displacing the sub-bands according to their lengths). Therefore, the solution proposed here is that once the sub-bands are ordered, only the coefficients placed in sub-bands longer than those where the parents are found are considered as their descendants. The remaining coefficients are coded separately. Table 2 shows the average bit-rate, SEGSNR, and PESQ results for some music, male and female speeches in 3 languages of English, German and Farsi. As seen in this Table, we can reduce the average bit-rates by about 10-25 % (as calculated on our test files) in comparison with the case where the sub-bands are displaced according to their lengths.

*2.3.1. SPIHT and Modified SPIHT*

Said and Pearlman introduced a new method named SPIHT for image coding [20]. The principles used in SPIHT are relatively the same as those in EZW. The Dominant and Subordinate passes in EZW are renamed as sorting and refinement in SPIHT, respectively. The refinement pass is the same as that used in EZW. The only difference is that the transmission order is shifted one level. That is, we do not have any output refinement bits in the first level with threshold $T_1$. However, the sorting pass alters as follows:

a) The following sets of coordinates are defined:

- $O(i,j)$: set of coordinates of all children of node (coefficient) $(i,j)$;
- $D(i,j)$: set of coordinates of all descendants of node $(i,j)$;
- $H$: set of coordinates of nodes in the highest pyramid level (lowest frequency sub-band);
- $L(i,j)$: $D(i,j) - O(i,j)$;

where $(i,j)$ is the pixel coordinates for images but, for audio signals they refer to coefficient coordinates in the time-frequency map.

b) The significance information is stored in three ordered lists, called list of insignificant sets (LIS), list of insignificant pixels (LIP), and list of significant pixels (LSP). In all lists each entry is identified by coordinate $(i,j)$, which in the LIP and LSP represents individual pixels, and in the LIS represents either the set $D(i,j)$ or $L(i,j)$. To differentiate between them, it is said that a LIS entry is of type A if it represents $D(i,j)$, and of type B if represents $L(i,j)$. Also, the following significance function is used to indicate the significance of a set of coordinates $\tau = \{(i,j)\}$ [20].

$$S_n(\tau) = \begin{cases} 1, & \max_{(i,j) \in \tau}\{|c_{i,j}|\} \geq 2^n \\ 0, & otherwise \end{cases} \quad (4)$$

where $2^n$ is equivalent to $T_\ell$ in EZW algorithm. The value $n$ is decremented by 1 in each next level. To simplify the notation of single coefficient sets, $S_n(\tau)$ is written as $S_n(i,j)$.

Then the sorting pass can be summarized as follows:

The coordinates of the lowest frequency sub-band coefficients are placed both in the LIS (as type A) and LIP. And, the LSP is left empty. The LIP entries are tested by function (4) and the result is outputted. If the coefficient is significant ($S_n(i,j) = 1$), its coordinate is moved to LSP and its sign is given as output. The LIS testing starts when the LIP testing is finished. If the entry $(i,j)$ is of type A, $S_n(D(i,j))$ is outputted. In the case of output "1", the coefficients of the set $O(i,j)$ are tested and acted upon in the same way as LIP entries. The coordinates of those insignificant coefficients are added to the LIP to be addressed in the next levels. In this case (output "1"), the entry is converted to type B and is moved to the end of the LIS. If the entry is of type B, $S_n(L(i,j))$ is outputted. If the result is "1", the entry $(i,j)$ is removed from the LIS and the coordinates of all coefficients in $O(i,j)$ are added to the end of the LIS as entries of type A and new tree roots. This process is continued to reach the end of the LIS. Then, the threshold is halved and all process is iterated. For all iterations, we will have the output refinement bits, as well. It is noted that the transmission of the refinement bits is started from the second level with threshold $T_2 = T_1/2$ (the first iteration). The algorithm is ended when it reaches the last step level with threshold 1 or satisfies the target bit-rate.

As stated in [31], the zero-tree (ZT) defined in EZW is simply a tree consisting of all zero values (named degree-0 ZT) while the zero-trees in SPIHT are defined in a wider sense. SPIHT can represent two more classes of zero-trees: 1-when all coefficients except parent are zero in a given tree (degree-1 ZT), and 2- when all coefficients except parent and its children are zero (degree-2 ZT). So, as shown in [20] and proved in [31], SPIHT can perform better than EZW for image compression. However, this comparison may not be valid for audio signals when wavelet packet is used and perceptual considerations are taken into account.

*Modified SPIHT*

Unlike EZW, the performance of SPIHT depends on the number of the sub-bands and the number of the children of each parent coefficient. In other words, the more the sub-bands and the less the children of each parent coefficient, the higher the output bits in the LIS checking part of the sorting pass. Furthermore, these factors increase also the computation time. As seen in Fig. 7, we have 26 sub-bands, as opposed to a dyadic case where the number of sub-bands is usually smaller than 10, and some coefficients may have only one child and therefore, it is not unexpected that the performance of the SPIHT algorithm worsens in our case.

To face this problem, we can define $O(i,j)$ in an ampler way than the original one. That is, in addition to direct descendants (children), we can also consider those in the next sub-bands. The experimental results show that the extension of $O(i,j)$ achieves better results, up to a given depth, and if this extension exceeds the bit-rate starts to increase. The depth $d$ means the $d$ sub-bands we have below the $i^{th}$ sub-band; for example depths 4 and 6 in Fig. 7. As used in Modified EZW, if we order the sub-bands according to the average magnitude of their coefficients, the probability of occurrence of high energy coefficients in the first sub-bands increases while this probability decreases as we move away from the first sub-bands. So, if we define $O(i,j)$ ampler in the first sub-bands and more restricted in the next ones, we can reach relatively suitable results.

Experimental results on the several different audio and speech signals show that better results are achievable if $O(i,j)$ is defined as follows:

- $O(i,j)$: set of coordinates of all descendants of coefficient $(i,j)$ in three descendant sub-bands after sub-band $i$ if $i = 1$ (the first sub-band) and otherwise, set of coordinates of all descendents in two descendant sub-bands after sub-band $i$.

The descendant sub-band means the sub-band with longer length than the sub-band where the parent coefficient is placed. If we apply these modifications together with those used in the Modified EZW, we can reach a 17-24 % reduction in the bit-rate compared to the SPIHT with displaced sub-bands on the basis of their lengths (Table 3(a-c)). To reach more reduction in bit-rate we exploit the scheme used in [32]. In that paper, reduced bit-rates have been obtained by merging the first and second sorting passes of the original SPIHT algorithm. That is, the sorting passes of the levels with $T_1$ and $T_1/2$ thresholds. As shown in Table 3(d), relatively better results are obtained with this modification. However, these results are almost the same as those of Table 2.

In some papers SPIHT is integrated, as such, in the audio or speech codec [33-37]. While, only in few reported works it is tried to modify and use it for audio or speech coding [28,38]. For example, in [33], where psycho-acoustic modeling and wavelet packet transform are used, the SPIHT algorithm is employed to encode the low frequency sub-bands while a reverse sorting process in addition with arithmetic coding is used to encode the high frequency ones. But, the disadvantage of this scheme resides in using arithmetic coding that, in addition to speed reduction, eliminates the possibility of truncating the output bit stream.

### 2.3.3. Proposed AVDZ algorithm

Our goal is to design a fast, efficient and scalable zero-tree-based algorithm that would outperform the two modified versions of EZW and SPIHT algorithms in terms of both coding efficiency and comparative computation times. To reach this goal we use the concept of degree-$k$ zero-trees as introduced in [31]. In that paper, the improvement of the coding performance of SPIHT for images is sought by defining the degree-$k$ zero-trees. But, in addition to increasing the complexity, no acceptable coding improvement is noted. We obtained same kind of results for audio signals. The degree-$k$ zero-tree is defined as a tree with all zero node values except in the top $k$ levels (top $k$ sub-bands in each tree in Fig. 7). On the other hand, the degree-$k$ zero-tree coder is defined as a source tree coder that can encode degree-$i$ zero-trees, $0 \leq i \leq k$. We use these definitions in another manner i.e., adaptively variable.

The proposed algorithm consists of two passes of refinement and sorting similar to SPIHT. The refinement pass is intact but the sorting is changed as follows. We have three lists named list of insignificant coefficients (LIC), list of significant coefficients (LSC) and list of insignificant sets (LIS) similar to SPIHT. But, unlike SPIHT where the LIS may grow during the coding process, this list has constant length (the number of the coordinates placed in it) equal to that of the first sub-band in our algorithm. In fact, only the coefficients of the first sub-band are considered as tree roots. The sub-bands are first ordered, according to the average magnitude of their coefficients, and then relocated as follows. By testing them, from top to bottom in Fig. 7, the first sub-band which has the minimum length (minimum number of coefficients) is moved up and placed as the first sub-band on the top of the time-frequency map. The other upper sub-bands are consequently moved down one location step. This is done for two reasons: 1- to solve the problem (c) mentioned in section 2.3.1. This problem can be tackled this way because there will be no shorter sub-band than the first one and hence the coefficients in other sub-bands can be considered as descendants of those in the first sub-band; 2- to minimize the length (the number of the coordinates as mentioned previously) of LIS and, thereby, decrease the output bit-rate and computation time because, for each coordinate in the LIS, the algorithm is to carry out some processing. The coordinates of the coefficients of the new first sub-band are placed both in LIS (as only tree roots) and LIC. The LSC is left empty. The LIC is dealt with in the same as LIP in SPIHT. But, for each coordinate in the LIS if all descendants are insignificant, "0" is outputted. Otherwise, the algorithm outputs "1" and moves down, in the current tree, from top to the sub-band after which there is no more significant descendant in comparison to the threshold in that level. This depth (i.e., the number of the sub-bands the algorithm moves down in each tree) is sent to the decoder as side information. The coordinates of coefficients found in this depth are added either to the LIC or LSC and are acted upon as $O(i,j)$ in SPIHT. No further action takes place in the sorting pass at this level. At the next level (with halved threshold) and after checking the coefficients of LIC, the algorithm continues the same process of testing the descendants for each coordinate of LIS from where it was left off in preceding level. In other words, during the coding process we have only *m* trees that their degrees may change adaptively in each level; *m* being the minimum sub-band length equal to that of the LIS.

Since we have ordered the sub-bands, the shorter depths are more probable to occur than longer ones and, we normally code these depths with only 2 bits. If the depth coding needs more than 2 bits, we first send a 2-bit zero code (which is unused in this part) and then carry the depth coding with 3 bits. We can continue this process up to 5 bits.

Finally, we can summarize the algorithm as follows:

1) **Ordering**:

    1.1) order the sub-bands according to the average magnitude of their coefficients, and send the necessary ordering sequence as side information;

    1.2) if the first sub-band is not the shortest do:

    - move up the first sub-band that is the shortest to the top of the time-frequency map;
    - move down the other upper sub-bands one step;

2) **Initialization**:

    2.1) set $n = \lfloor \log_2(\max_{(i,j)}\{|c_{i,j}|\}) \rfloor$ and $S_n(\tau)$ as equation (4);

    2.2) output $n + 1$ as the last step level;

    2.3) set the LSC as an empty list;

    2.4) set $k_j\big|_{j=1}^{L_1} = 1$ ($L_1$ is the length of sub-band 1 which is now the shortest and $k_j$ is the degree of the zero-tree whose root is found in coordinate $(1, j)$);

    2.5) add the coordinates $(1, j)$ to the LIC and $(1, j, k_j)$ to the LIS;

3) **Sorting Pass**:

    3.1) for each entry $(i, j)$ in the LIC do:

    - output $S_n(i, j)$;
    - if $S_n(i, j) = 1$ then move $(i, j)$ to the LSC and output the sign of $c_{i,j}$;

    3.2) for each entry $(1, j, k_j)$ in the LIS do:

    - if $k_j$ is not equal to the last sub-band number (26 in our work) then:
        - $\tau$ = set of coordinates of all descendants of coefficient $(1, j)$ from sub-band $(k_j + 1)$ to the last;
        - Output $S_n(\tau)$;
        - if $S_n(\tau) = 1$ then:
            - $k_{j(old)} = k_j$;
            - move down, in current tree, from sub-band $(k_j + 1)$ to sub-band after which there is no more significant descendant in comparison to the current threshold $2^n$;
            - $k_j = k_{j(new)}$ = the new sub-band number;
            - $depth = k_{j(new)} - k_{j(old)}$ and send it as side information;

- for each $(a,b)$ in the calculated $depth$ do:
  - output $S_n(a,b)$;
  - if $S_n(a,b) = 1$ then add $(a,b)$ to the LSC and output the sign of $c_{a,b}$;
  - if $S_n(a,b) = 0$ then add $(a,b)$ to the LIC;

4) **Refinement Pass**: the same as SPIHT;

5) **Decision for the next level**: if $n \neq 0$ decrease $n$ by 1 and go to step 3.

Figure 9 shows the adaptive variable degree-$k$ zero-trees for the first two levels of the proposed AVDZ algorithm. The sub-bands, in this figure, are ordered according to the order of Fig. 8(d). As it is seen, the first sub-band (from left to right in Fig. 8(d) or up to down in Fig. 9) that is the shortest (sub-band 7 before ordering in this case) is moved up and placed first on the top of the time-frequency map after ordering. This way, the problems (c) and (d) are solved simultaneously, as mentioned previously. Therefore, we will expect that the AVDZ algorithm outperforms the modified EZW and SPIHT algorithms.

Perceptually quantized WPT coefficients of an input frame may be all zero (e.g., in silent parts). In these cases, the last step level in part 2.2 of the algorithm is sent as zero and no more information is needed to be sent to the decoder.

The decoding algorithm is straight forward and can be obtained by simply reversing the coding process. In other words, when the decoder receives data, the three control lists (LIC, LSC, and LIS) are formed identical to those used in the encoder. Therefore, it is easily seen that coding and decoding have the same computational complexity in this scheme. The major task done by the decoder is to reconstruct the signal gradually, level by level. For the value $n$ when a coordinate is moved to the LSC, it is known that $2^n \leq |c_{i,j}| \leq 2^{n+1}$. So, the decoder uses this information in addition to the received sign bit, just after the insertion of the coordinate in the LSC, to set the reconstructed value as $\pm 1.5 \times 2^n$. Since the sending of the refinement bits is shifted one level, both in our algorithm and SPIHT, in comparison to EZW, the decoder adds to or subtracts from the magnitude of the reconstructed coefficient value the amount $2^{n-1}$ ($= T_\ell/2$ instead of $T_\ell/4$ in EZW) when it receives the refinement bits of "1" or "0", respectively. In this manner, the distortion decreases gradually during both the sorting and refinement passes.

As seen in Table 4, this algorithm can reduce the average bit-rates by about 6-10% in comparison with both the modified EZW and SPIHT algorithms. It is noted that the results of the EZW and SPIHT algorithms are obtained by displacing the sub-bands according to their lengths. The results obtained for the Modified EZW and SPIHT algorithms are the best of those shown in Tables 2 and 3, respectively. Table 5 shows the coding and decoding computation times for our signal examples.

All reported computation times, in this paper, refer to the same computer of 2.8 GHz CPU and 1.25 GB of RAM. As seen in this table, the introduced algorithm has a very shorter coding and decoding computation times in comparison with the modified EZW and SPIHT schemes. Comparisons using more examples are given in the following.

## 3. Results

To more accurately compare the used zero-tree-based schemes and to better evaluate the proposed AVDZ algorithm, we use 24 different male and female speech signals in 3 languages of English, German and Farsi (Fig. 10). The first and second letters of each file name specify the language and gender, respectively. All these files have an equal time length of 4 seconds, and as mentioned previously are sampled at 16 kHz sampling frequency and represented with 16 bit/sample. These speeches include also those used in the previously given tables. The same kind of comparative results are obtained for music signals but are not included in this figure because of their high bit-rates that would jeopardize the presentations. Figure 11 shows the average coding, decoding and total computation times, averaged on 24 speech files, for each method. The total time is the sum of the coding and decoding computation times. As seen in this figure, the proposed algorithm is considerably superior to all others, by about 35% and 50% relative to Modified SPIHT and EZW algorithms, respectively. Both perceptual (in PESQ scores) and objective (in SEGSNR (dB)) measures of the reconstructed signals, indicated in Fig. 12, show the good performance of the codec. Moreover, the good resulting perceptual qualities could be considered as a sign of the correct performance of the proposed perceptual quantizer of section 2.2. To evaluate the correct performance of our proposed perceptual quantizer, we compare it with the case of quantizing all coefficients with 7 bits, referred to as 7-bit uniform quantizer. In both of these codecs, the AVDZ algorithm is used as re-encoder. As seen in Fig. 13, the proposed perceptual quantizer can reduce the average bit-rate by about 15-35% with almost the same or even better perceptual quality.

As stated before, the zero-tree-based methods have the desirable property of generating an embedded binary code of desired size. Truncation of their output binary stream does not produce perceivable artifacts since it only eliminates the least significant refinement bits of the encoded coefficient values. Since we have ordered the sub-bands for each input frame, the coefficients with higher magnitudes are sent earlier, than that is usually done in conventional EZW and SPIHT, with some beneficial effect in bit stream truncation. Fig. 14 indicates the average, minimum and maximum PESQ scores for the proposed AVDZ algorithm in the fixed bit-rates of 24, 32, 40 and 48 kbit/sec. The averaging is computed on about 500 male and female random-selected TIMIT files with different dialects. To avoid excessive time delay between coder and decoder, we use only buffers of one frame long (32 msec) for bit-rate control. As seen in this figure, the proposed algorithm achieves respectively, on average, 3.6, 4, 4.2 and 4.3 PESQ scores. Figure 15, reported from [39], shows the perceptual qualities (in MOS scores) of the wideband ITU-T standard codecs G.722, G.722.1 and G.722.2. As concluded in [39], the subjective MOS scores of these codecs are strongly correlated with their corresponding objective MOS values (e.g., PESQ). However, slight differences may be seen between the results of different codecs. For G.722 and G.722.2, the PESQ scores are slightly lower than the MOS scores [39,40]. However, by comparing the Figs. 14 and 15, it seems that the proposed codec yields better perceptual qualities than the wideband standard codecs G.722 and G.722.1.

## 4. Conclusion

In this paper we introduce a new wideband speech and audio codec. The coder consists of: wavelet packet transform, perceptual quantizer, and zero-tree-based re-encoder. The problems of exploiting two well-known EZW and SPIHT algorithms, as re-encoder, are studied and discussed. Modifications are proposed for these two algorithms that improve their efficiencies by about 15-25%. Moreover, a new algorithm coined "adaptive variable degree-$k$ zero-tree (AVDZ)" is introduced that outperforms, both in terms of average bit-rate and speed, the modified versions of EZW and SPIHT re-encoders. In addition, it has the beneficial properties of generating embedded fixed bit-rates and equal coding and decoding computation times as with other zero-tree-based codecs. A simple perceptual quantizer is also incorporated that reduces the average bit-rate by about 15-35%, with almost the same or better perceptual quality, when used in conjunction with the new AVDZ algorithm, in comparison to the case where no perceptual considerations are used. Finally, it is concluded that the introduced codec can achieve, on average, the PESQ scores of 3.6, 4, and 4.2 for the bit-rates of 24, 32, and 40 kbit/sec which seem to be better than those reported for wideband ITU-T standards G.722 and G.722.1. We intend to expand this research, in the future, using the statistical properties of the non-zero values of WPT coefficients and more efficient perceptual quantizers.

**Figures and Tables**

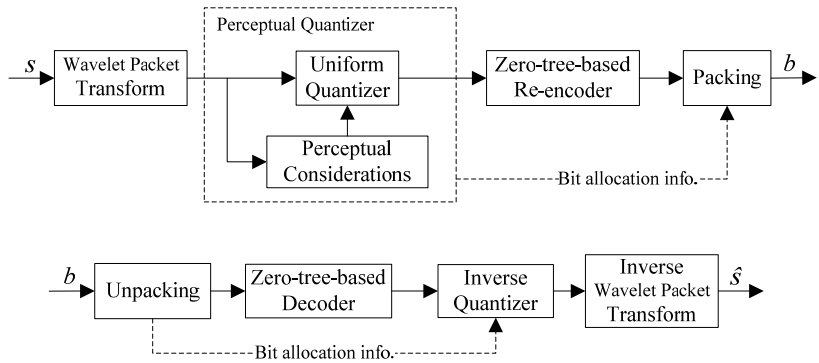

Fig. 1.  Block diagram of the coder (top) and decoder (bottom).

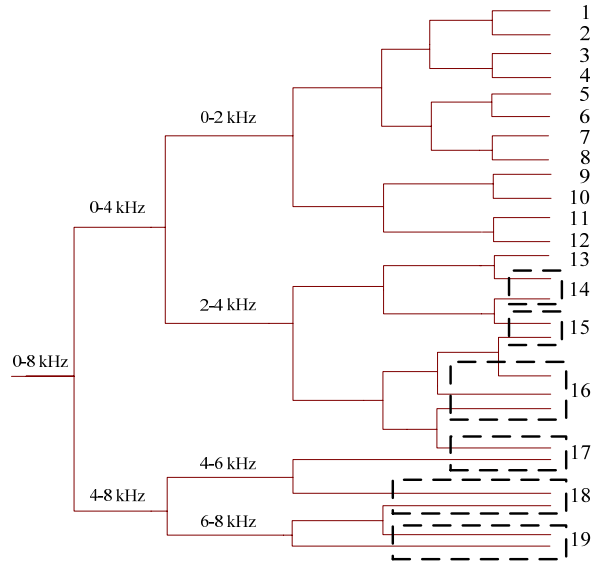

Fig. 2.  Decomposing the input frame according to critical sub-bands.

Table 1  Nineteen critical sub-bands used

| Band Number | Frequency Range (Hz) |
| --- | --- |
| 1 | 0-125 |
| 2 | 125-250 |
| 3 | 250-375 |
| 4 | 375-500 |
| 5 | 500-625 |
| 6 | 625-750 |
| 7 | 750-875 |
| 8 | 875-1000 |
| 9 | 1000-1250 |
| 10 | 1250-1500 |
| 11 | 1500-1750 |
| 12 | 1750-2000 |
| 13 | 2000-2250 |
| 14 | 2250-(2500)-2750 |
| 15 | 2750-(3000)-3125 |
| 16 | 3125-(3250-3500)-3750 |
| 17 | 3750-(4000)-5000 |
| 18 | 5000-(6000)-6500 |
| 19 | 6500-(7000)-8000 |

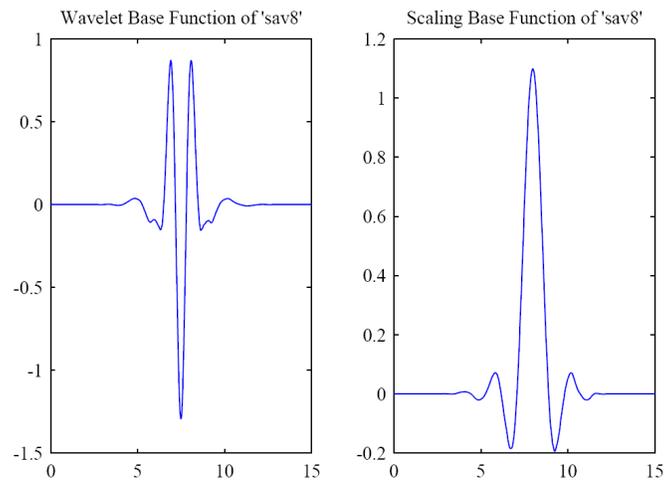

Fig. 3.  The wavelet and scaling base functions of "sav8" with $\beta = 0.168$.

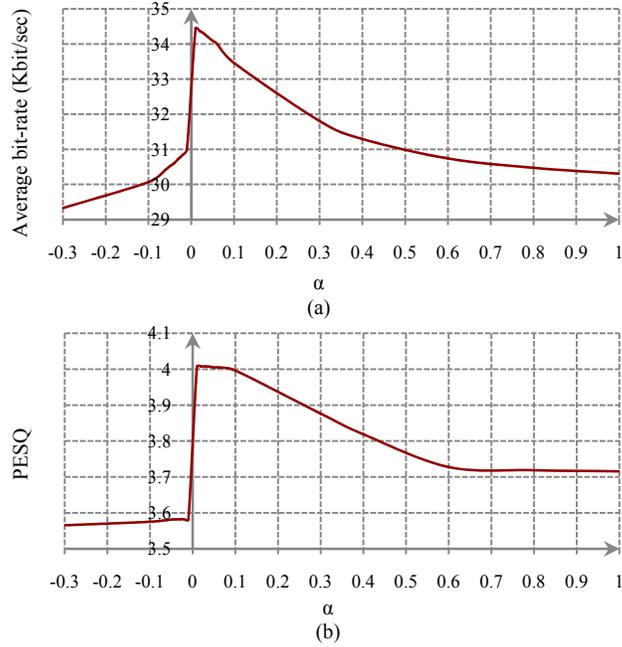

Fig. 4. Determining the factor $\alpha$ for a randomly selected signal, (a) average bit-rates, and (b) their corresponding perceptual qualities versus $\alpha$.

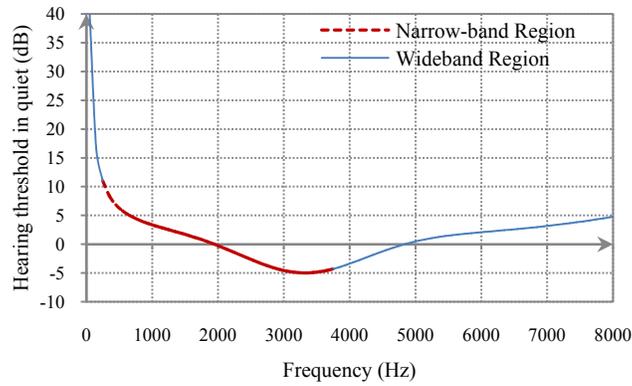

Fig. 5. Comparison of the hearing sensitivity of the human ear in the wideband and narrow-band regions according to our classification.

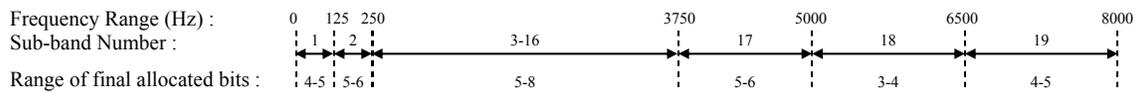

Fig. 6. Range of final allocated bits for each sub-band.

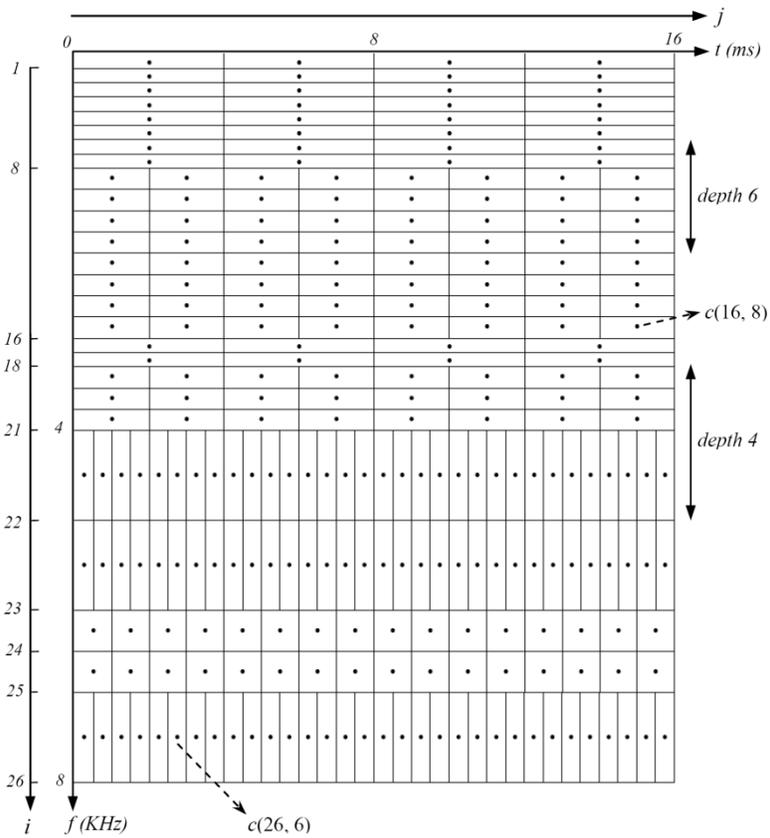

Fig. 7. Time-Frequency map according to our decomposition for a 256-sample input frame. And illustrating how the coefficient $c(i, j)$ and depth $d$ are defined.

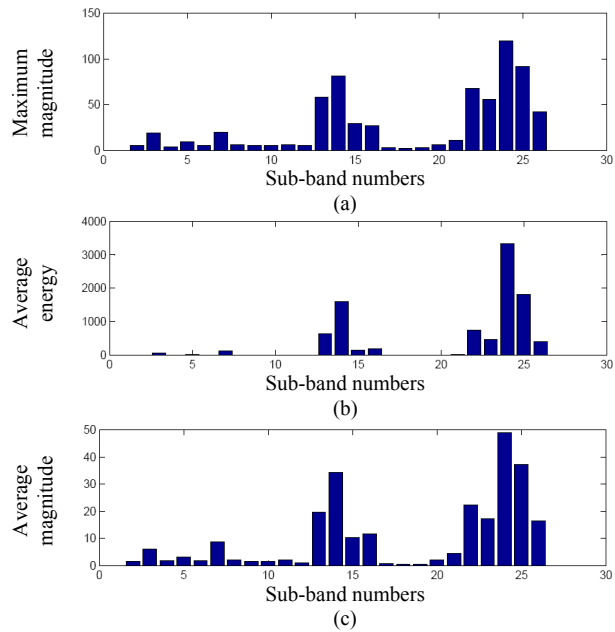

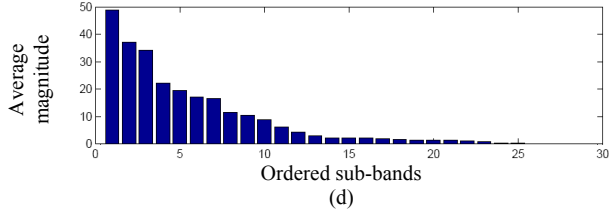

Fig. 8. Ordering the sub-bands of a typical input frame, before ordering (a-c), after ordering (d).

Table 2 Results of EZW scheme, (a) relocating the sub-bands according to their lengths, (b) ordering the sub-bands on the basis of the average magnitude of their coefficients (speeches with suffix numbers 1 to 3 are in Farsi, German and English languages, respectively)

| Test signal | Length (sec) | SEGSNR (dB) | PESQ | Average bit-rate (bit/sec) | |
|---|---|---|---|---|---|
| | | | | (a) | (b) |
| Male 1 | 4 | 23.91 | 4.20 | 39945 | 34994 |
| Female 1 | 4 | 21.57 | 4.17 | 39269 | 35900 |
| Male 2 | 4 | 20.36 | 3.93 | 38346 | 32095 |
| Female 2 | 4 | 19.68 | 3.93 | 37955 | 31850 |
| Male 3 | 4 | 25.05 | 3.86 | 40788 | 30659 |
| Female 3 | 4 | 25.33 | 3.81 | 41690 | 31861 |
| Music 1 | 4 | 18.51 | 4.42 | 60020 | 53181 |
| Music 2 | 4 | 22.49 | 4.42 | 63284 | 50872 |

Table 3 Results of SPIHT scheme, (a) displacing the sub-bands in the basis of their length, (b) ordering the sub-bands based on the average magnitude of their coefficients, (c) modification of b based on modified $O(i, j)$, (d) modification of c based on merging sorting passes of the first two levels

| Test signal | Average bit-rate (bit/sec) | | | |
|---|---|---|---|---|
| | (a) | (b) | (c) | (d) |
| Male 1 | 43784 | 38477 | 34564 | 34418 |
| Female 1 | 43391 | 39138 | 35342 | 35255 |
| Male 2 | 41374 | 35603 | 32203 | 31753 |
| Female 2 | 41583 | 35131 | 31831 | 31592 |
| Male 3 | 41614 | 35399 | 32606 | 31782 |
| Female 3 | 42929 | 36367 | 33364 | 32813 |
| Music 1 | 63841 | 58154 | 52898 | 52615 |
| Music 2 | 65727 | 57270 | 53261 | 51497 |

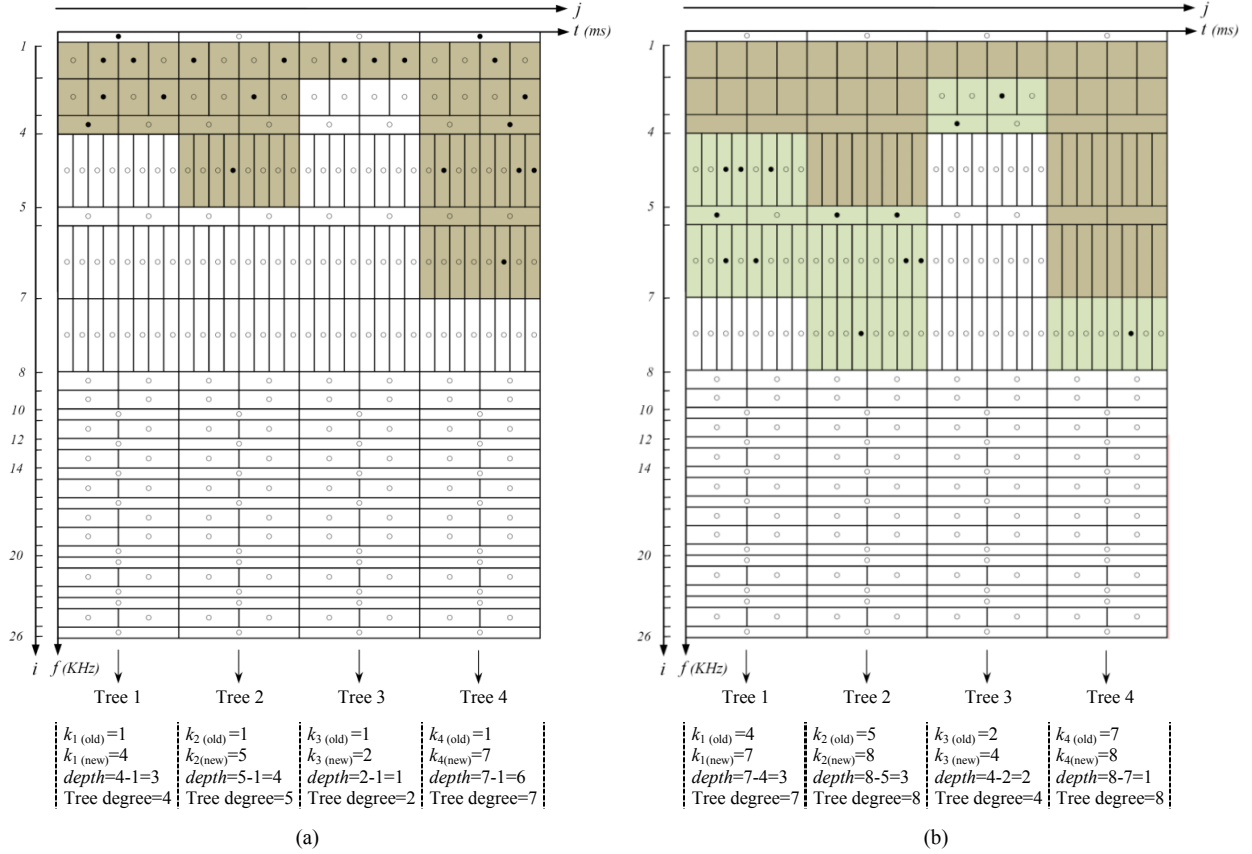

Fig. 9. The performance of the proposed AVDZ algorithm in sorting passes of (a) level 1, and (b) level 2 for an assumed input frame with sub-band ordering of Fig. 8(d). The solid and hollow circles represent respectively the significant and insignificant coefficients relative to the corresponding thresholds of those levels.

Table 4  Comparison of the re-encoder average bit-rates

| Test signal | SEGSNR (dB) | PESQ | Average bit-rate (bit/sec) | | | | |
| --- | --- | --- | --- | --- | --- | --- | --- |
| | | | EZW | SPIHT | Modified EZW | Modified SPIHT | Proposed AVDZ |
| Male 1 | 23.91 | 4.20 | 39945 | 43784 | 34994 | 34418 | 33159 |
| Female 1 | 21.57 | 4.17 | 39269 | 43391 | 35900 | 35255 | 33521 |
| Male 2 | 20.36 | 3.93 | 38346 | 41374 | 32095 | 31753 | 29921 |
| Female 2 | 19.68 | 3.93 | 37955 | 41583 | 31850 | 31592 | 29689 |
| Male 3 | 25.05 | 3.86 | 40788 | 41614 | 30659 | 31782 | 28749 |
| Female 3 | 25.33 | 3.81 | 41690 | 42929 | 31861 | 32813 | 29853 |
| Music 1 | 18.51 | 4.42 | 60020 | 63841 | 53181 | 52615 | 48503 |
| Music 2 | 22.49 | 4.42 | 63284 | 65727 | 50872 | 51497 | 46530 |

Table 5 Comparison of the re-encoder computation times

| Test signal | Coding computation time (sec) | | | Decoding computation time (sec) | | |
| --- | --- | --- | --- | --- | --- | --- |
| | Modified EZW | Modified SPIHT | Proposed AVDZ | Modified EZW | Modified SPIHT | Proposed AVDZ |
| Male 1 | 25.35 | 18.25 | 11.41 | 21.34 | 17.67 | 11.85 |
| Female 1 | 25.58 | 18.62 | 10.80 | 22.71 | 17.49 | 12.17 |
| Male 2 | 26.38 | 18.81 | 11.13 | 21.74 | 17.08 | 12.50 |
| Female 2 | 26.34 | 17.32 | 11.14 | 21.13 | 17.46 | 12.56 |
| Male 3 | 25.91 | 17.29 | 10.97 | 19.29 | 16.51 | 12.41 |
| Female 3 | 23.26 | 17.48 | 11.14 | 21.33 | 16.30 | 11.62 |
| Music 1 | 30.81 | 22.62 | 11.79 | 24.49 | 20.16 | 13.07 |
| Music 2 | 33.44 | 22.13 | 11.74 | 25.80 | 19.54 | 13.36 |

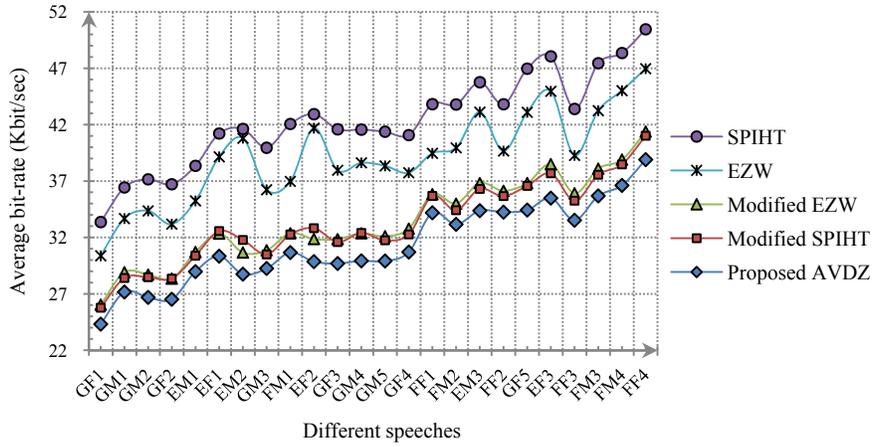

Fig. 10. Comparison of the average bit-rates of the re-encoders for 24 different Male and Female speeches in 3 languages of English, German and Farsi (The first and second letters of each speech name introduce its language and gender, respectively).

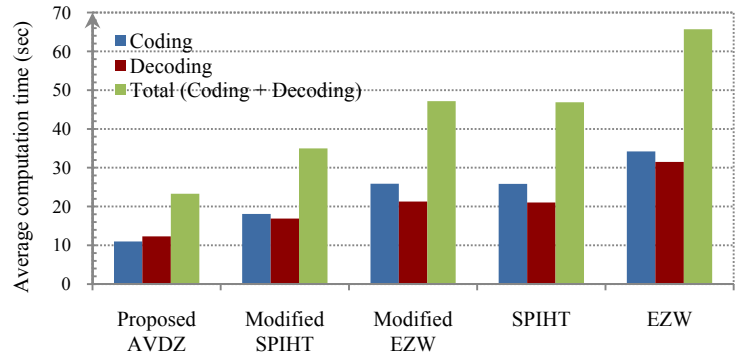

Fig. 11. Comparison of the average coding, decoding and total computation times.

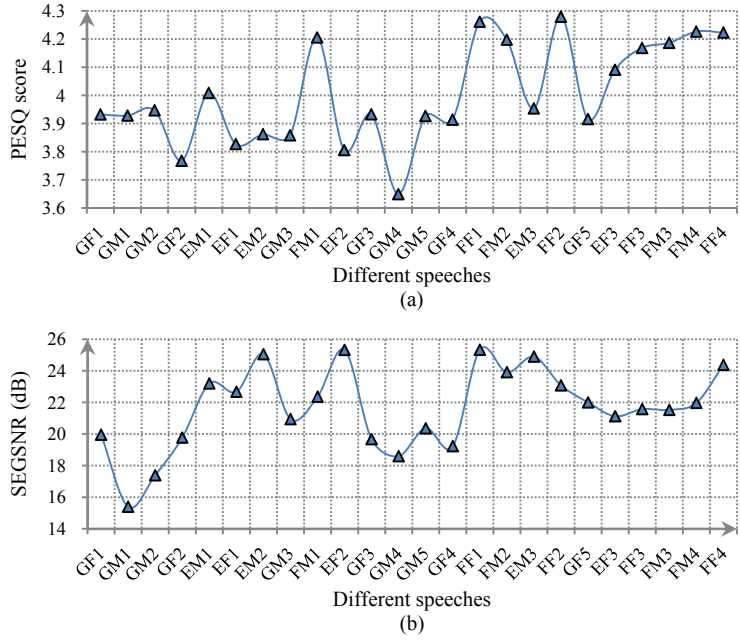

Fig. 12. Perceptual (a) and objective (b) measures of the reconstructed signals.

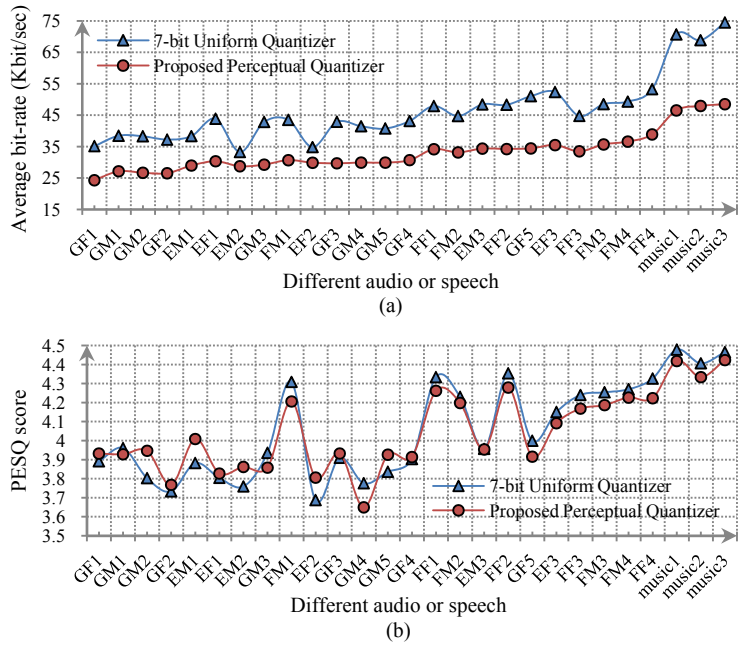

Fig. 13. Comparison of the proposed perceptual quantizer with a 7-bit uniform quantizer.

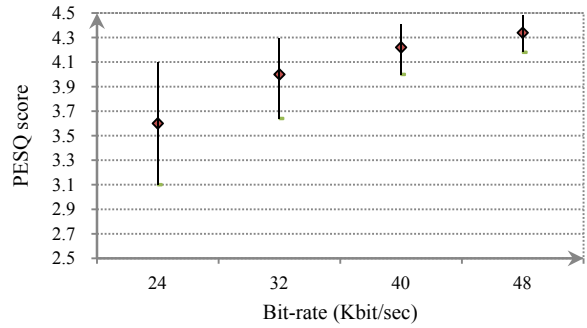

Fig. 14. Average, minimum and maximum PESQ scores for the introduced wideband codec, computed on about 500 male and female random-selected TIMIT files with different dialects.

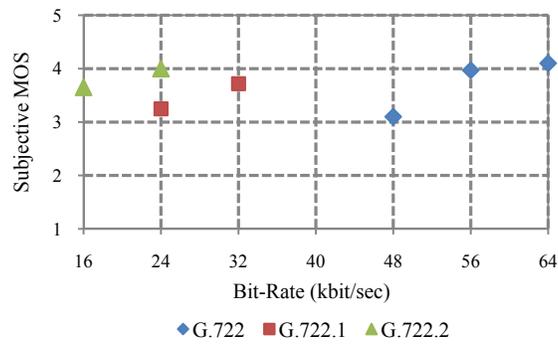

Fig. 15. Subjective MOS scores for ITU-T wideband standard codecs [39].

30